# Enhanced flexoelectricity at reduced dimensions revealed by mechanically tunable quantum tunnelling

Saikat Das[1,2], Bo Wang[3], Tula R. Paudel[4], Sung Min Park[1,2], Evgeny Y. Tsymbal[4], Long-Qing Chen[3], Daesu Lee[5]* and Tae Won Noh[1,2]*

[1]Center for Correlated Electron Systems, Institute for Basic Science (IBS), Seoul 08826, Korea.

[2]Department of Physics and Astronomy, Seoul National University, Seoul 08826, Korea.

[3]Department of Materials Science and Engineering, Pennsylvania State University, University Park, Pennsylvania 16802, USA.

[4]Department of Physics and Astronomy & Nebraska Center for Materials and Nanoscience, University of Nebraska, Lincoln, Nebraska 68588, USA.

[5]Department of Physics, Pohang University of Science and Technology (POSTECH), Pohang 37673, Korea

Correspondence and requests for materials should be addressed to D.L. (email: dlee1@postech.ac.kr) and T.W.N. (email: twnoh@snu.ac.kr).

**Abstract**

**Flexoelectricity is a universal electromechanical coupling effect whereby all dielectric materials polarize in response to strain gradients. In particular, nanoscale flexoelectricity promises exotic phenomena and functions, but reliable characterization methods are required to unlock its potential. Here, we report anomalous mechanical control of quantum tunnelling that allows for characterizing nanoscale flexoelectricity. By applying strain gradients with an atomic force microscope tip, we systematically polarize an ultrathin film of otherwise nonpolar $SrTiO_3$, and simultaneously measure tunnel current across it. The measured tunnel current exhibits critical behaviour as a function of strain gradients, which manifests large modification of tunnel barrier profiles via flexoelectricity. Further analysis of this critical behaviour reveals significantly enhanced flexocoupling strength in ultrathin $SrTiO_3$, compared to that in bulk, rendering flexoelectricity more potent at the nanoscale. Our study not only suggests possible applications exploiting dynamic mechanical control of quantum effect, but also paves the way to characterize nanoscale flexoelectricity.**

**Introduction**

Polar materials form the basis of electromechanics, optoelectronics, and studies on emerging quantum states[1]. Such materials belong to only 10 of the 32 possible crystal point groups, and sometimes exhibit problematic size effects[2]. Under such circumstances, flexoelectricity[3–5] offers unique advantages[6–15]. Strain gradients can intrinsically polarize all materials with arbitrary crystal symmetries[3–5], ranging from dielectrics[16] to semiconductors[11] and from bio-materials[17] to two-dimensional materials. Importantly, such ubiquitous flexoelectric effects potentially become even larger at the nanoscale, as strain gradients scale inversely with material size. Nanoscale strain-graded dielectrics (e.g., a strain variation $\Delta u =$ 1% within 1 nm) encompass enormous strain gradients (to $\partial u/\partial x = 10^7$ m$^{-1}$) and may exhibit remarkable phenomena and flexoelectric functionality[6–15]. Furthermore, nanoscale flexoelectricity can fundamentally differ from the conventional bulk flexoelectricity, e.g., due to a nonlinear polarization response under large strain gradients[9]. Thus, characterizing nanoscale flexoelectricity, namely, the flexocoupling coefficient, is of great importance from both a fundamental and technological viewpoint.

To characterize the flexocoupling coefficients at the nanoscale, it is necessary to identify a nanoscale phenomenon that can be actively controlled by the flexoelctric effect, which could allow quantifying the flexoelectric polarization as a function of strain-gradient. It is well established that the quantum tunnelling probability through a nanometer thick ferroelectric barrier layer sandwiched between two metallic electrodes sensitively depends on the polarization direction and its magnitude[18–20]. In this so-called ferroelectric tunnel device, the depolarization field, originating from the imperfect screening of ferroelectric polarization by the metallic electrodes, alters the intrinsic barrier height. For asymmetric electrodes, changing the polarization direction yields two different effective barrier heights, and subsequently leads to two discrete electroresistance states. Meanwhile, due to the converse piezoelectric effect, the barrier width can also modulate in response to the electric field applied during the tunnelling transport measurement[21–25]. This also leads to dissimilar electroresistance states. All these considerations, suggest a possibility of controlling quantum tunnelling by the flexoelectric effect, thereby quantifying the flexocoupling coefficient at the nanoscale.

Here, we demonstrate that a systematic control of quantum tunnelling through a paraelectric ultrathin SrTiO$_3$ (STO) film by strain-gradient induced flexoelectric polarization allows quantifying flexocoupling coefficient at the nanoscale. By applying the strain-gradients from a conductive scanning probe tip we simultaneously polarize and measure the tunnelling current across the film. With increasing strain-gradients, the tunnelling current exhibits an asymmetric-symmetric crossover, which we attribute, based on the Wentzel–Kramers–Brillouin (WKB) modelling, to flexoelectric polarization-driven modification of the tunneling barrier profile. Furthermore, analyzing the modification of the barrier profile as a function of strain-gradients enables quantifying the flexocoupling coefficient, which we found to be significantly enhanced compared to the bulk. We discuss possible origins of this enhanced flexocoupling coefficient.

**Results**

**Concept of flexoelectric control of quantum tunnelling**

Figure 1a shows a schematic of our experimental setup. We used a conductive atomic force microscope (AFM) tip (PtIr-coated) to apply strain gradients[8] and simultaneously measure the tunnel current. We systematically generated giant strain gradients (up to $> 10^7$ m$^{-1}$), as estimated by contact mechanics analysis (Fig. 1b and Methods). These strain gradients are much larger than those achievable using a conventional beam-bending approach, which generates gradients in the range of $10^{-1}$ m$^{-1}$ (using micrometre-thick beams)[16] to $10^2$ m$^{-1}$ (employing nanometre-thick beams)[10]. When an ultrathin dielectric layer becomes flexoelectrically polarized by a giant strain gradient, the resulting depolarization field and electrostatic contribution[18–20] significantly modify the tunnel barrier profile (Figs. 2a–c). Therefore, we can utilize pure mechanical force by an AFM tip as a dynamic tool not only for systematically controlling quantum tunnelling, but also for characterizing nanoscale flexoelectricity.

As a model system, we chose the archetypal dielectric material SrTiO$_3$ (STO), which remains paraelectric down to a temperature of 0 K in bulk. We prepared homoepitaxial ultrathin STO films on (001)-oriented STO substrates with a conductive SrRuO$_3$ (SRO) buffer layer. To avoid the off-stoichiometry-driven ferroelectric phase of STO [26], we used an ultra-slow growth scheme[27], combined with in situ post-annealing in oxygen to minimize oxygen vacancies.

Piezoresponse force microscopy confirmed that the STO films are indeed paraelectric (Supplementary Fig. 1). Notably, our geometry induced compressive strains in both the transverse $x_1$ and longitudinal $x_3$ directions (Fig. 1b and Supplementary Fig. 4), attributable to AFM tip-induced downward bending and pressing. Such three-dimensional compression of STO does not favour either ferroelectricity or piezoelectricity. This makes it possible to explore pure flexoelectric polarization, the electrostatic effect of which modifies the STO tunnel barrier[18,19].

**Strain-gradient dependent tunnelling transport**

We measured the tunnel current across a nine unit cell-thick (i.e., ~3.5 nm-thick) STO as a function of the applied strain gradients. Our theoretical analysis reveals that the transverse strain gradients defined as,

$$\partial u_t/\partial x_3 \ (= \partial u_{11}/\partial x_3 + \partial u_{22}/\partial x_3) \tag{1}$$

are an order of magnitude larger than the longitudinal strain gradients $\partial u_{33}/\partial x_3$ (Supplementary Note 2 and Supplementary Fig. 4); we thus considered only the $\partial u_t/\partial x_3$ values. Figures 2d–f show the measured current–voltage (*I–V*) curves for three representative $\partial u_t/\partial x_3$ values (see Supplementary Fig. 2 for the entire set). For $\partial u_t/\partial x_3 < 1.56 \times 10^7 \ \mathrm{m^{-1}}$ (Fig. 2d), the *I–V* curves exhibit typical tunnelling characteristics (red solid line) and are highly asymmetric, manifesting rectifying behaviour. The forward current systematically increases with increasing $\partial u_t/\partial x_3$, whereas the reverse current remains comparable to the noise level (~1 pA). When $\partial u_t/\partial x_3$ attains a critical value, $1.56 \times 10^7 \ \mathrm{m^{-1}}$, the *I–V* curve became maximally asymmetric (Fig. 2e). Beyond a $\partial u_t/\partial x_3$ of $1.56 \times 10^7 \ \mathrm{m^{-1}}$, however, the reverse current begins to increase, whereas the forward current increased only marginally, rendering the *I–V* curve more symmetric (Fig. 2f). Figure 2g emphasizes this critical behaviour by plotting rectification ratios ($RR \equiv |I_{+V}/I_{-V}|$) as a function of $\partial u_t/\partial x_3$. We also observed a similar critical behaviour in the eleven unit-cell thick STO film (Supplementary Fig. 3).

Before addressing how flexoelectricity could explain these results, we rule out other possible origins of the phenomena. First, the AFM tip-induced pressure did not cause any permanent surface damage to the STO film (Supplementary Fig. 5). Additionally, the mechanical control of electron tunnelling is reversible (Supplementary Fig. 6), excluding any involvement of an electrochemical process. We also considered the effect of strain on the STO tunnel barrier profiles. AFM tip-induced compressive strain per se would not only decrease the barrier width ($\Delta d \leq 0.2$ nm) but also slightly increase the STO band gap[28] and hence the barrier height. However, our detailed analysis show that the strain effect is too small to explain our observations (Supplementary Note 3). Furthermore, we confirmed that the strain-induced changes in electronic properties of SRO are too small to be responsible for the anomalous behaviour of tunnelling transport (Supplementary Note 6). Thus, the asymmetric-symmetric crossover is an intrinsic effect possibly attributable to flexoelectric polarization-induced modification of the tunnel barrier.

**Understanding and modelling the tunnelling transport**

To understand how the barrier profile affects tunnel current, we perform a one-dimensional Wentzel–Kramers–Brillouin (WKB) simulation of a metal-insulator-metal (M1-I-M2) heterostructure (Supplementary Note 4). In the experiment, 'M1', 'M2', and 'I' correspond to SRO, PtIr, and STO, respectively. Our calculations suggest that the observed rectifying tunnelling behaviour should originate from an asymmetric, trapezoidal barrier profile, with the barrier height $\varphi_1$ at the M1-I interface being smaller than the barrier height $\varphi_2$ at the I-M2 interface (as in Fig. 2a). Such an asymmetric tunnel barrier implies downward flexoelectric polarization (pointing towards the M1/I interface), and a higher probability of transmission to the M1 electrode than in the reverse direction (to the M2 electrode). When flexoelectric polarization attains a critical value, the tunnel barrier becomes triangular, such that $\varphi_1 = 0$ (as in Fig. 2b), yielding the maximum rectifying behaviour. This explains the anomalous increase in *RR* in regime (A) (yellow) of Fig. 2h.

When the flexoelectric polarization increases further, the conduction band minimum of STO could cross the Fermi level (as in Fig. 2c). This crossing metallizes the interfacial barrier layer and concomitantly decreases the effective barrier width *d*, as supported by first-principles calculations (Fig. 3, Supplementary Note 5, and Methods). For convenience, we describe this case using a negative $\varphi_1$ (Supplementary Fig. 10). As the *RR* of a triangular tunnel barrier is exponentially proportional to the barrier width *d*, the decrease in *d* would lower the *RR*, as shown in regime (B) (blue) of Fig. 2h. Notably, the barrier-asymmetry dependence of *RR* (Fig. 2h) strikingly resembles the experimentally observed strain-gradient dependence of *RR* (Fig. 2g); both exhibit the asymmetric-symmetric crossover. Therefore, we conclude that flexoelectric polarization-induced metallization near the SRO/STO interface manifests itself as an asymmetric-symmetric crossover in tunnelling transport.

Next, to understand how the barrier profile varies with increasing $\partial u_t/\partial x_3$, we fit the tunnel spectra of regime (A) to an analytical equation[20] that describes tunnelling through a trapezoidal barrier (red solid line in Fig. 2d; see Methods and Supplementary Note 1). Taking into account the work functions of SRO (5.2 eV)[29] and PtIr (5.6 eV)[30], and the electron affinity of STO (3.9 eV)[29], we set the intrinsic barrier heights $\varphi_{0,1}$ and $\varphi_{0,2}$ to 1.3 and 1.7 eV, respectively (black line in Fig. 2a). Furthermore, following simple electrostatics argument[18,19], we constrain the $\varphi_1$ and $\varphi_2$ to vary obeying the relation: $\Delta\varphi_1/\Delta\varphi_2 = (\varphi_{0,1} - \varphi_1)/(\varphi_2 - \varphi_{0,2}) = \delta_{SRO}/\delta_{PtIr}$, where $\delta_{SRO}$ and $\delta_{PtIr}$ are the effective screening lengths of SRO and PtIr, respectively. Given that $\delta_{SRO} \approx 0.5$–0.6 nm [2] and $\delta_{PtIr} < 0.1$ nm, we set $\Delta\varphi_1/\Delta\varphi_2$ (= $\delta_{SRO}/\delta_{PtIr}$) to be 8. Figure 4a plots the fitted $\varphi_1$ and $\varphi_2$ values as a function of $\partial u_t/\partial x_3$. Consistent with the WKB simulations, our fitting yields highly asymmetric trapezoidal barrier profiles, where with increasing $\partial u_t/\partial x_3$, $\varphi_1$ decreases from 0.57 to 0.34 eV, and $\varphi_2$ increases from 1.79 to 1.82 eV.

**Quantifying flexocoupling coefficient at the nanoscale**

Based on this $\partial u_t/\partial x_3$ dependence of $\varphi_1$ and $\varphi_2$, we estimate the strength of effective flexoelectric coupling. The transverse strain gradient polarizes the STO layer through the flexoelectric effect, and the induced polarization can be expressed as $P = \varepsilon \cdot f_{\text{eff}} \cdot (\partial u_t/\partial x_3)$, where $\varepsilon$ and $f_{\text{eff}}$ are the dielectric permittivity and the effective flexocoupling coefficient of STO, respectively. In ultrathin STO, this flexoelectric polarization results in depolarization field ($\propto -P/\varepsilon$) and modifies the tunnel barrier profile according to the following electrostatic equation (Supplementary Note 7)[18,19]:

$$(\varphi_2 - \varphi_1)/ed = P/\varepsilon + E_{\text{bi}} = f_{\text{eff}} \cdot (\partial u_t/\partial x_3) + E_{\text{bi}}, \qquad (2)$$

where $e$ is the electronic charge and $E_{\text{bi}}$ is the additional built-in field contribution that could arise from the work function difference between SRO and PtIr, surface dipoles[31], and/or an offset between the calculated and actual strain gradients. As shown in Fig. 4b, the calculated $(\varphi_2 - \varphi_1)/ed$ varies almost linearly with $\partial u_t/\partial x_3$ (grey solid line), giving a slope $f_{\text{eff}}$ of 23 ± 1 V. In addition, fitting also yielded a nonzero contribution at $\partial u_t/\partial x_3 = 0$ (i.e., 8–10×10$^7$ V m$^{-1}$), corresponding to the built-in field $E_{\text{bi}}$.

We now focus on the onset of asymmetric-symmetric crossover of tunnel current at $(\partial u_t/\partial x_3)_c = 1.56 \times 10^7$ m$^{-1}$, which also allows us to estimate $f_{\text{eff}}$. According to our simulation, this crossover is attributable to the polarization-induced trapezoidal-to-triangular transition of the tunnel barrier. At the critical $\partial u_t/\partial x_3$ (or equivalently, at the critical $P$), we therefore expect that $\varphi_1 = 0$ and $\varphi_2 = \varphi_{0,2} + \varphi_{0,1} \cdot (\delta_{\text{PtIr}}/\delta_{\text{SRO}}) = 1.7 + 1.3/8$ eV, giving $(\varphi_2 - \varphi_1)/ed = 5.32 \times 10^8$ V m$^{-1}$ (Fig. 4). With $E_{\text{bi}} = 9 \times 10^7$ V m$^{-1}$ and $(\partial u_t/\partial x_3)_c = 1.56 \times 10^7$ m$^{-1}$, equation (2) yields $f_{\text{eff}}$

= 28 V. This value compares reasonably well to that (23 ± 1 V) obtained from fitting, demonstrating that our approach is innately consistent. Furthermore, using this $f_{\text{eff}}$, we simulated a three-dimensional profile of local $P$ (Fig. 1c and Methods), and found that the average out-of-plane $P$ was around 0.17 C m$^{-2}$ for $\partial u_t/\partial x_3 = 1.6 \times 10^7$ m$^{-1}$ (i.e., just above the critical $\partial u_t/\partial x_3$). This value compares well to the predicted critical polarization (i.e., $P_c = 0.16$ C m$^{-2}$; Supplementary Note 8), which again emphasizes the self-consistency of our approach.

**Discussion**

Interestingly, the estimated flexocoupling coefficient (23–28 V) is larger than Kogan's phenomenological estimate (1–10 V)[3,5], and indeed an order of magnitude greater than the experimental value (~2.6 V) for bulk STO[16]. To understand this enhancement, we first note that a nonlinear flexoelectric response could arise under large strain gradients, as demonstrated in several material systems[9,32]. By considering the nonlinear flexoelectricity, e.g., third-order response (Supplementary Note 9), we might explain the enhancement of $f_{\text{eff}}$ under a huge $\partial u_t/\partial x_3$. Additionally, a surface contribution $f_{\text{surf}}$ can be involved[33,34], which, combined with the bulk contribution $f_{\text{bulk}}$, determines the overall coupling coefficient $f_{\text{eff}}$ (= $f_{\text{surf}}$ + $f_{\text{bulk}}$) of a material. When considered separately, both $f_{\text{surf}}$ and $f_{\text{bulk}}$ could be >10 V in magnitude but opposite in sign[35]. Our results may thus imply that only either the surface or bulk contribution becomes dominant in the ultrathin limit. To obtain a complete understanding of enhanced flexoelectricity in ultrathin STO, further systematic experimental and theoretical investigations will be required.

In summary, we show that quantum tunnelling is mechanically tunable. Such mechanical tunability allows determining the flexocoupling strength at the nanoscale, which we found to be much enhanced compared to that in bulk. This finding emphasizes that flexoelectricity could become much more powerful at reduced dimensions due to not only a large strain gradient, but also an enhanced coupling strength. We hope that this study would encourage the construction of flexocoupling coefficient databases at the nanoscale, and the design of high-performance flexoelectric devices. From a broader perspective, this study highlights several favourable aspects of nanoscale flexoelectricity. First, nanoscale flexoelectricity allows for the generation of large polarization in a continuous manner. We started from a nonpolar STO, and continuously polarized it with $P$ up to 0.4 C m$^{-2}$. Second, such a continuously tunable, large polarization can also generate a large electrostatic potential, which corresponds to a stationary effective electric field, as high as $10^9$ V m$^{-1}$. This can be useful for a large electric-field control of dielectrics, which has been challenging due to dielectric breakdown.

**Methods**

**Sample fabrication.** SRO and STO thin films were sequentially grown on $TiO_2$-terminated and (100)-oriented STO substrates. The growth dynamics and thicknesses were monitored by in situ reflection high-energy electron diffraction (RHEED). Film deposition was performed at 700°C under oxygen partial pressures of 100 and 7 mTorr for SRO and STO, respectively. After deposition, films were annealed at 475°C for 1 h in oxygen at ambient pressure and subsequently cooled to room temperature at 50°C min$^{-1}$. Structural characterization, namely, the reciprocal space mapping was performed to ensure that the STO film is strain-free (Supplementary Fig. 14).

**Tunnelling measurements.** Current-voltage curves were obtained using an Asylum Research Cypher AFM at room temperature under ambient conditions. Conducting PtIr-coated metallic tips (NANOSENSORS™ PPP-EFM) with nominal spring constants 50–60 N m$^{-1}$, and a dual-gain ORCA module, were used to measure currents. An electrical bias was applied through the conducting SRO electrode; this was swiped from −1 V to +1 V at a ramping rate of about 4 V s$^{-1}$. The noise floor of the AFM system was about ~1 pA.

To extract barrier heights from the tunnelling *I–V* curves, we used an analytical equation describing direct tunnelling through trapezoidal tunnel barriers[20,36]:

$$I(V) \cong b + c \frac{\exp\left\{\alpha(V)\left[(\varphi_2 - \frac{eV}{2})^{\frac{3}{2}} - (\varphi_1 + \frac{eV}{2})^{\frac{3}{2}}\right]\right\}}{\alpha^2(V)\left[(\varphi_2 - \frac{eV}{2})^{\frac{1}{2}} - (\varphi_1 + \frac{eV}{2})^{\frac{1}{2}}\right]^2} \sinh\left\{\frac{3}{2}\alpha(V)\left[(\varphi_2 - \frac{eV}{2})^{\frac{1}{2}} - (\varphi_1 + \frac{eV}{2})^{\frac{1}{2}}\right]\frac{eV}{2}\right\}. \quad (3)$$

where *c* is a constant and $\alpha(V) \equiv [4d(2m_e)^{1/2}]/[3\hbar(\varphi_1 + eV - \varphi_2)]$. Also, *b*, $m_e$, *d*, and $\varphi_{1,2}$ are the baseline, free electron mass, barrier width, and barrier height, respectively. As explained in the main text, our fittings imposed the constraints $\varphi_2 = 1.7 + \Delta\varphi$ and $\varphi_1 = 1.3 - 8\Delta\varphi$. In addition, we used a scaling factor to account for the increase in contact area with increasing contact force, but this did not affect our principal results (i.e., the *RRs*, $|I_{+V}/I_{-V}|$). For smaller $\Delta\varphi$ values, we used the entire tunnelling spectra for fitting (Supplementary Figs. 2a–c). However, when larger distortions of the barrier profiles were apparent (i.e., at larger $\Delta\varphi$ values), we fitted the tunnelling spectra using smaller bias windows.

**Simulation of strain profile.** The strain distribution in a 3.5 nm-thick STO thin film pressed with an AFM tip is obtained by solving the elastic equilibrium equation by using Khachaturyan microelasticity theory[37] and the Stroh formalism of anisotropic elasticity[38]. The detailed procedure has been elaborated in previous works[39]. Here, we discretized three-dimensional space into 64 × 64 × 700 grid points and applied periodic boundary conditions along the $x_1$ and $x_2$ axes. The grid spacing was $\Delta x_1 = \Delta x_2 = 1$ nm and $\Delta x_3 = 0.1$ nm. Along the $x_3$ direction, 35 layers were used to mimic the film; the relaxation depth of the substrate featured 640 layers to ensure that the displacement at the bottom of substrate was negligibly small. To estimate surface stress distribution that developed on AFM-tip pressing, we adopted the Hertz contact mechanics of the spherical indenter[40] with a tip radius of 30 nm and a mechanical force of 1–7 μN. The Young's moduli and Poisson ratios of the Pt tip and the STO film were $E^{\text{tip}} = 168$ GPa and $\upsilon^{\text{tip}} = 0.38$, and $E^{\text{film}} = 264$ GPa and $\upsilon^{\text{film}} = 0.24$, adapted from reference 41. The electrostrictive and rotostrictive coupling coefficients of STO were adapted from reference 42. See Supplementary Note 2 for more details.

**Simulation of polarization profile.** The polarization distribution under the mechanical load by an AFM tip can be calculated by self-consistent phase-field modeling[43]. The temporal evolution of polarization field $\mathbf{P}(x,t)$ is governed by the time-dependent Ginzburg-Landau equation, i.e., $\partial\mathbf{P}/\partial t = -L(\delta F(\mathbf{P})/\delta \mathbf{P})$, where $L$ is the kinetic coefficient. The total free energy $F$ can be expressed as[43]

$$F = \int (f_{bulk} + f_{elastic} + f_{electric} + f_{gradient} + f_{flexo}) dV$$

$$= \int [\alpha_{ij} P_i P_j + \alpha_{ijkl} P_i P_j P_k P_l + \beta_{ij} \theta_i \theta_j + \beta_{ijkl} \theta_i \theta_j \theta_k \theta_l + t_{ijkl} P_i P_j \theta_k \theta_l + \frac{1}{2} g_{ijkl} \frac{\partial P_i}{\partial x_j} \frac{\partial P_k}{\partial x_l} \quad (4)$$

$$+ \frac{1}{2} k_{ijkl} \frac{\partial \theta_i}{\partial x_j} \frac{\partial \theta_k}{\partial x_l} + \frac{1}{2} c_{ijkl} (\varepsilon_{ij} - \varepsilon_{ij}^0)(\varepsilon_{kl} - \varepsilon_{kl}^0) - \frac{1}{2} E_i P_i + \frac{1}{2} f_{ijkl} (\frac{\partial P_k}{\partial x_l} \varepsilon_{ij} - \frac{\partial \varepsilon_{ij}}{\partial x_l} P_k)] dV$$

The bulk Landau free energy $f_{bulk}$ consists of two sets of order parameters, i.e., the spontaneous polarization **P** and the antiferrodistortive order parameter **θ**, which represents the oxygen octahedral rotation angle of STO[42]. The flexoelectric contribution is considered as a Liftshitz invariant term as

$$f_{flexo} = \frac{1}{2} f_{ijkl} (\frac{\partial P_k}{\partial x_l} \varepsilon_{ij} - \frac{\partial \varepsilon_{ij}}{\partial x_l} P_k). \quad (5)$$

The eigenstrain tensor $\varepsilon^0$ in the elastic energy density is given by

$$\varepsilon_{ij}^0 = Q_{ijkl} P_k P_l + \Lambda_{ijkl} \theta_k \theta_l - F_{ijkl} P_{k,l}, \quad (6)$$

where the electrostrictive, rotostrictive, and converse flexoelectric couplings are considered via tensor **Q**, **Λ** and **F**. The coefficients used in constructing the total free energy $F$ of STO single crystal were given in our previous works[42,44]. The transverse flexoelectric constant of STO estimated from experiments in present work were used ($f_{12}$ = 25 V) while the other two flexoelectric component are assumed to be zero (i.e., $f_{11} = f_{44} = 0$) for simplicity.

**First-principles calculations.** The atomic and electronic structure of the system was obtained using the density functional theory (DFT) implemented in the Vienna *ab initio* simulation package (VASP)[45,46]. The projected augmented plane wave (PAW) method was used to approximate the electron-ion potential[47]. The exchange and correlation potentials were calculated using the local spin density approximation (LSDA). In calculation, we employed a kinetic energy cutoff of 340 eV for PAW expansion, and a 6 × 6 × 1 grid of **k** points[48] for Brillouin zone integration. The in-plane lattice constant was that of relaxed bulk STO ($a$ = 3.86 Å); the $c/a$ ratio and the internal atomic coordinates were relaxed until the Hellman–Feynman force on each atom fell below |0.01| eV Å$^{-1}$. The dielectric constant were calculated using density functional perturbation theory[49–51]. See Supplementary Note 5 for more details.

**Data availability**

All relevant data presented in this manuscript are available from the authors upon reasonable request.

**Acknowledgements**

This work was supported by the Research Center Program of the IBS (Institute for Basic Science) in Korea (grant no. IBS-R009-D1). D.L. acknowledges the support by the National Research Foundation of Korea (NRF) grant funded by the Korea government (MSIT) (No. 2018R1A5A6075964). B.W. acknowledges the support by the NSF-MRSEC grant number DMR-1420620. The effort of L.Q.C. is supported by National Science Foundation (NSF) through Grant No. DMR-1744213. The work at the Pennsylvania State University used the Extreme Science and Engineering Discovery Environment (XSEDE) Bridges at the Pittsburgh Supercomputing Center through allocation TG-DMR170006, which is supported by National Science Foundation grant number ACI-1548562[52]. The research at the University of Nebraska−Lincoln is supported by the National Science Foundation through the Nebraska Materials Science and Engineering Center (MRSEC), Grant No. DMR-1420645.

**Author Contributions**

D.L. conceived and designed the research. D.L. and T.W.N. directed the project. S.D. fabricated thin films and measured tunnelling transport assisted by S.M.P. B.W. carried out simulations of strain gradient and polarization under the supervision of L.-Q.C. T.R.P. and E.Y.T. carried out first principles calculations. S.D. and D.L. prepared the manuscript. All authors discussed the results and implications and commented on the manuscript at all stages.

**Competing Interests statement**

The authors declare no competing interests.

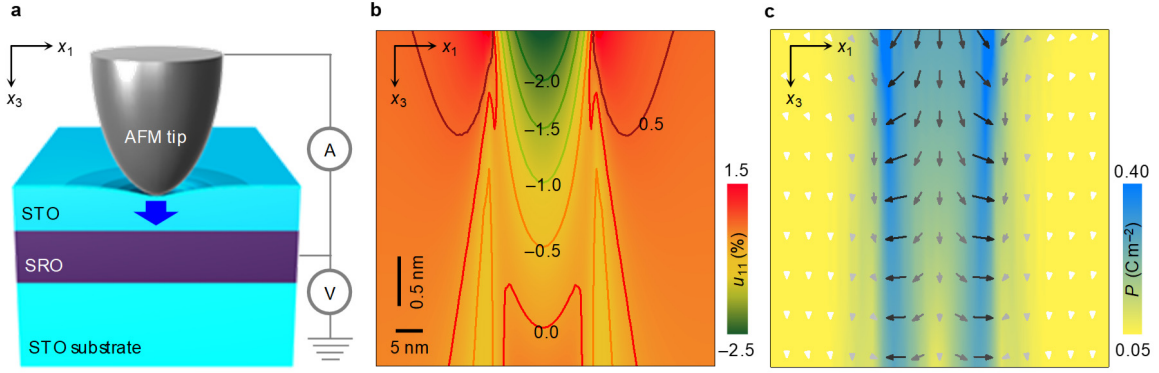

**Fig. 1. Electron tunnelling through a flexoelectrically polarized ultrathin barrier. a**, Schematic of the experimental setup, illustrating flexoelectric polarization (blue arrow) generated by the atomic force microscope (AFM) tip pressing the surface of ultrathin $SrTiO_3$ (STO). **b**, Simulated transverse strain $u_{11}$ in a nine unit cell-thick (i.e., 3.5 nm-thick) STO under a representative tip loading force of 5 μN. Along the central line, $u_{11}$ varies by ~0.5% within $\Delta x_3$ = 0.5 nm, yielding $\partial u_{11}/\partial x_3 \sim 10^7$ m$^{-1}$. **c**, Polarization profile, obtained by phase-field simulation, for the strain profile in **b**. Arrows denote the polarization direction. In the tip-contact region, the polarization along the $x_3$ direction was around 0.17 C m$^{-2}$ on average. Note that when neglecting flexoelectricity (i.e., $f = 0$), our simulation does not produce any polarization in STO, which again confirms the flexoelectricity-based origin of our observation.

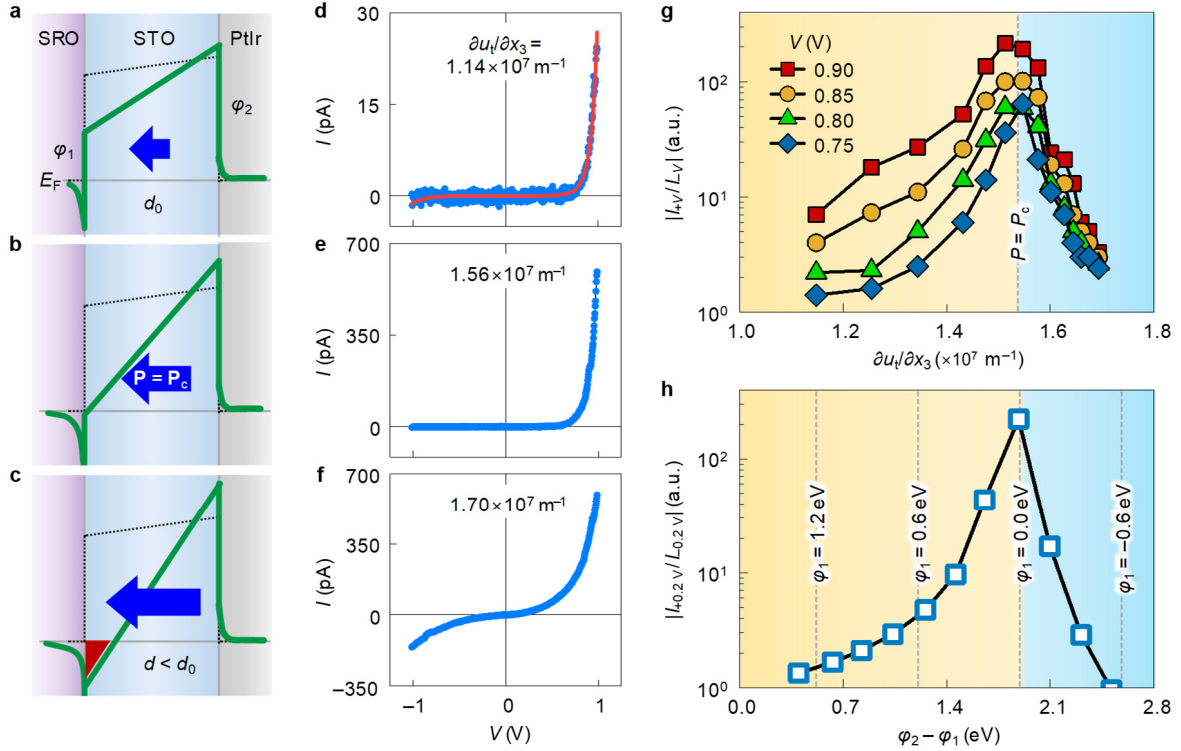

**Fig. 2. Flexoelectric control of electron tunnelling. a–c**, Schematics of the potential energy profiles across SrTiO$_3$ (STO) with increasing flexoelectric polarization (*P*; blue arrow*s*). The additional electrostatic potential induced by *P* modifies the original barrier potential energy (black line) to yield the total potential energy (solid green line). At the critical polarization $P_c$, the tunnel barrier becomes triangular with $\varphi_1 = 0$ and $\varphi_2 = \varphi_{0,2} + \varphi_{0,1} \cdot (\delta_{PtIr}/\delta_{SRO})$. **d–f**, Measured tunnel current–voltage (*I–V*) curves across the nine unit cell-thick STO film for three representative $\partial u_t/\partial x_3$ values. The red solid line in **d** indicates the it to Equation (3). **g**, The rectification ratios (*RRs*) $|I_{+V}/I_{-V}|$ of the measured tunnel current as a function of $\partial u_t/\partial x_3$. With increasing $\partial u_t/\partial x_3$, the tunnelling *I–V* curves become more asymmetric in regime (A) (yellow) below $\partial u_t/\partial x_3 = 1.56 \times 10^7$ m$^{-1}$, but more symmetric in regime (B) (blue). **h**, The simulated $|I_{+0.2V}/I_{-0.2V}|$ at *V* = 0.2 V as a function of barrier asymmetry, defined as $\varphi_2 - \varphi_1$. We set the initial barrier heights as $\varphi_1 = 1.3$ eV and $\varphi_2 = 1.7$ eV, and systematically decreased $\varphi_1$ (or increased $\varphi_2$) while fulfilling the condition $(1.3 - \varphi_1)/(\varphi_2 - 1.7) = 8$.

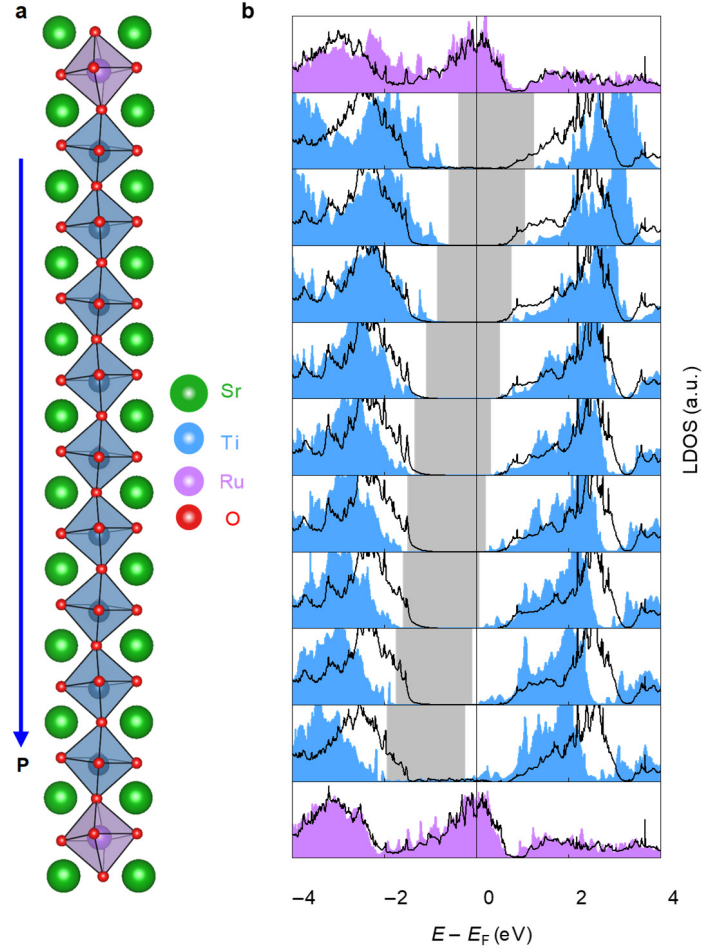

**Fig. 3. Polarization-induced local metallization in SrTiO₃. a**, The simulation cell. We artificially polarized SrTiO₃ (STO) layers with uniform displacement of Ti atom by 0.2 Å. **b**, Calculated layer-resolved density of states (LDOS) of polarized STO layers (filled blue) compared to that of nonpolar STO (black solid line). Grey regions represent a gap between conduction band minimum and valence band maximum of polarized STO layers, clearly showing a shift of energy bands due to polarization-induced electric field.

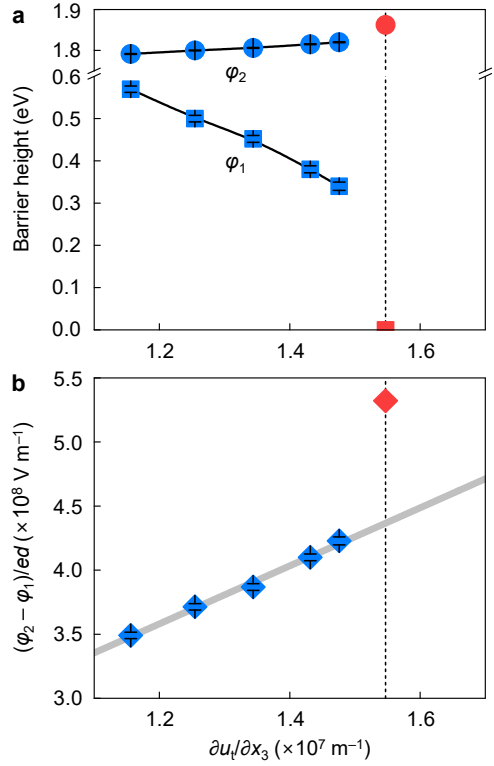

**Fig. 4. Characterizing flexoelectricity in ultrathin SrTiO$_3$. a**, The blue squares and circles indicate $\varphi_1$ and $\varphi_2$, respectively, extracted by fitting the tunnelling spectra to the Eq. 3. The error bars represent the standard errors of the extracted barrier heights.. The red square and circle represent $\varphi_1$ and $\varphi_2$, respectively, for the triangular barrier at the critical $\partial u_t/\partial x_3$. **b**, $(\varphi_2 - \varphi_1)/ed$, calculated from **a**. The grey line shows a linear fit to the data. The error bars represent the standard errors.

**Supplementary Information**

**Enhanced flexoelectricity at reduced dimensions revealed by mechanically tunable quantum tunnelling**

Saikat Das et al.

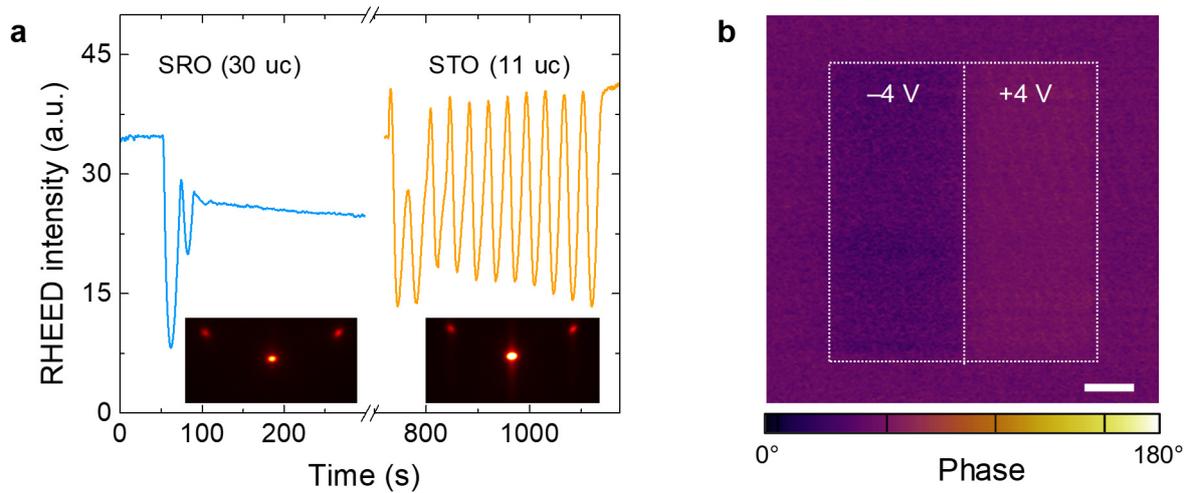

**Supplementary Figure 1. Growth and electrical characterization.** **a,** In situ growth monitoring employing reflection high-energy diffraction (RHEED). The growth mode of the metallic SrRuO$_3$ (SRO) layer exhibited the typical transition from layer-by-layer to step-flow, as evident by the evolution of RHEED intensity (blue curve). The SrTiO$_3$ (STO) layer, however, grew in a layer-by-layer manner. Here, we show representative RHEED intensity profiles collected during the growth of 30 unit cell-thick SRO and 11 unit cell-thick STO layers. The RHEED patterns obtained after growth (shown underneath the intensity profiles) feature sharp Bragg reflexes, indicative of atomically smooth surfaces. **b,** The piezoresponse force microscopy (PFM) phase image of an eleven unit cell-thick STO film taken after electrical poling with ±4V of applied bias. No phase contrast is discernible across the poled area, suggesting that our STO film was paraelectric. We obtained similar PFM phase images for the nine- and five-unit cell-thick films.

**Supplementary Note 1**

To extract barrier heights from the tunneling I–V curves, we used an analytical equation describing direct tunneling through trapezoidal tunnel barriers[1,2]:

$$I(V) \cong b + c \frac{\exp\left\{\alpha(V)\left[(\varphi_2 - \frac{eV}{2})^{\frac{3}{2}} - (\varphi_1 + \frac{eV}{2})^{\frac{3}{2}}\right]\right\}}{\alpha^2(V)\left[(\varphi_2 - \frac{eV}{2})^{\frac{1}{2}} - (\varphi_1 + \frac{eV}{2})^{\frac{1}{2}}\right]^2} \sinh\left\{\frac{3}{2}\alpha(V)\left[(\varphi_2 - \frac{eV}{2})^{\frac{1}{2}} - (\varphi_1 + \frac{eV}{2})^{\frac{1}{2}}\right]\frac{eV}{2}\right\}, \quad (1)$$

where $c$ is a constant and $\alpha(V) \equiv [4d(2m_e)^{1/2}]/[3\hbar(\varphi_1 + eV - \varphi_2)]$. Also, $b$, $m_e$, $d$, and $\varphi_{1,2}$ are the baseline, the electron mass, barrier width, and barrier height, respectively.

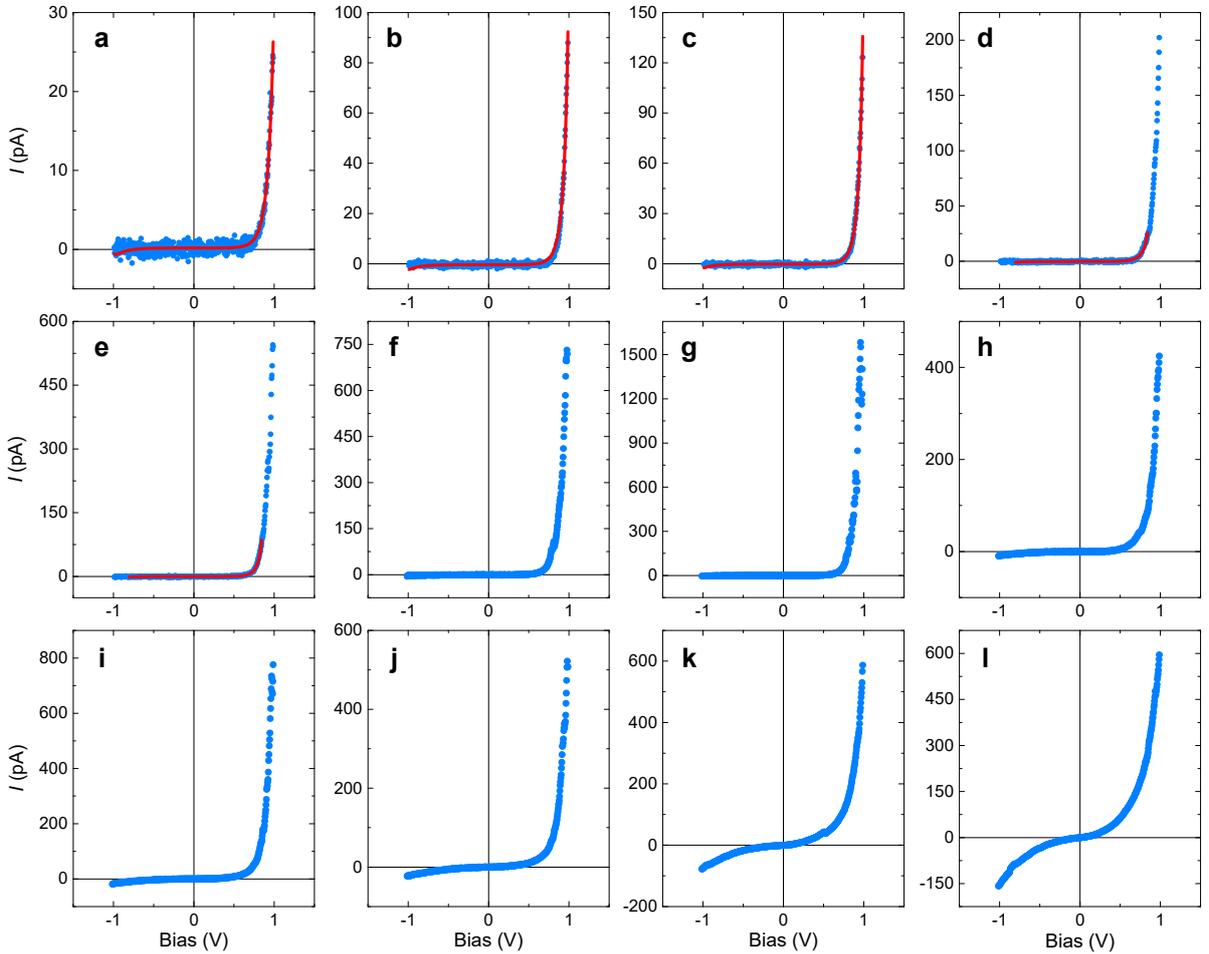

**Supplementary Figure 2. Tunneling currents across a nine-unit cell-thick SrTiO₃ film with increasing strain gradients.** Tunneling currents were measured as the applied strain gradient increased from $1.14 \times 10^7$ m$^{-1}$ (**a**) to $1.7 \times 10^7$ m$^{-1}$ (**l**). Figure 2 of the main text presents spectra **a**, **f**, and **l**. The solid red lines in Figures **a**–**e** indicate the fits to Supplementary Eq. 1. Note that in **a**–**c**, we fitted the entire spectra (i.e., –1 V to +1 V), but we used smaller bias windows to fit the tunneling currents of **d** and **f**.

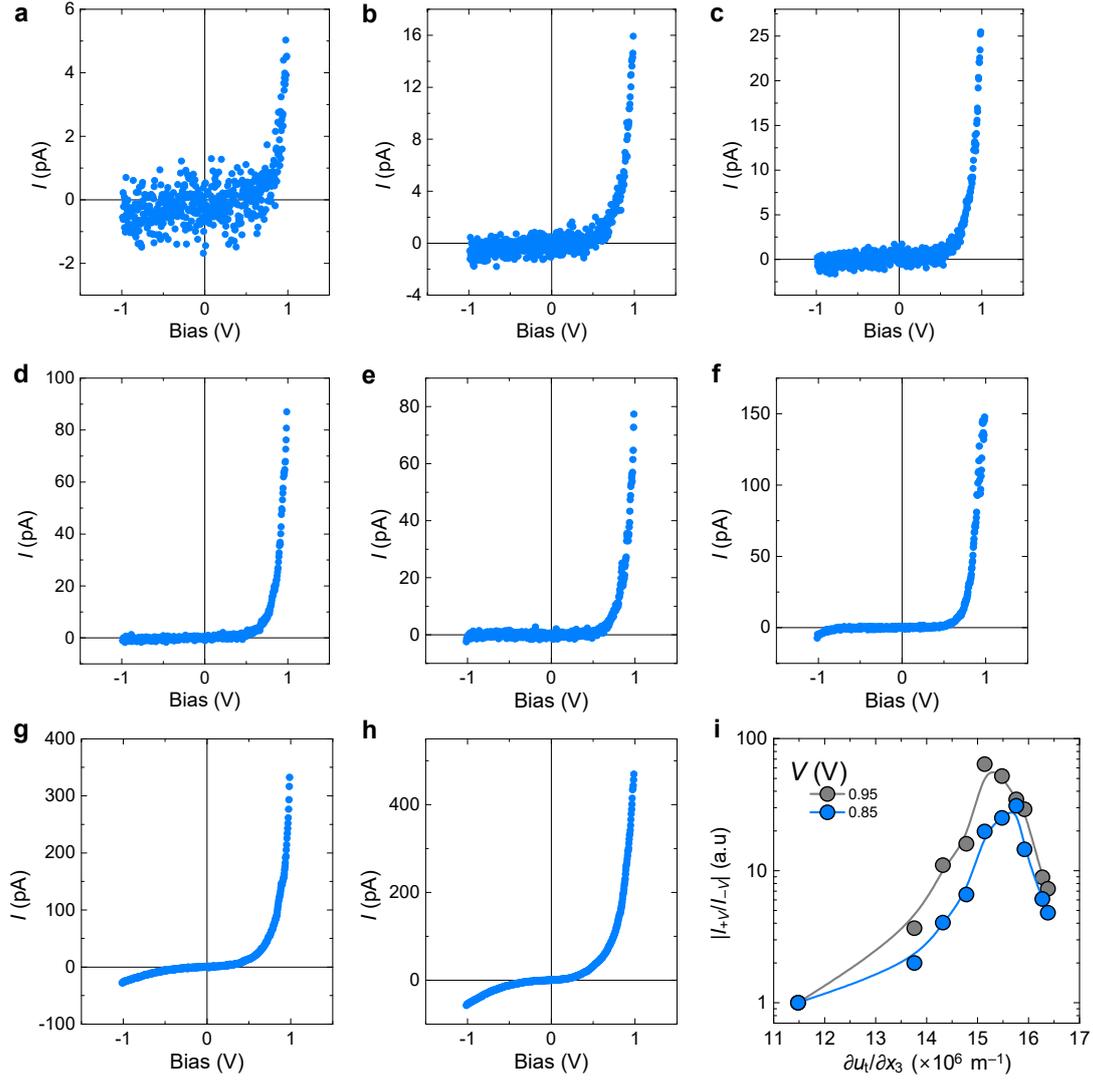

**Supplementary Figure 3. Tunneling currents across an eleven-unit cell-thick SrTiO$_3$ film**. **a–h**, Tunneling currents measured as the applied strain gradient increased from $1.15 \times 10^7$ m$^{-1}$ (**a**) to $1.63 \times 10^7$ m$^{-1}$ (**h**). **i**, *RRs*, |$I_{+V}/I_{-V}$|, plotted as a function of $\partial\varepsilon_t/\partial x_3$, showing the asymmetric-symmetric crossover. We used two biases, 0.95 and 0.85 V, for calculation.

**Supplementary Note 2**

We considered an STO thin film of thickness $h_f$, with the top surface in contact with an AFM tip and the bottom interface coherently constrained by the substrate. At the top surface, the normal stress distribution (as a function of the distance from the contact center) is described by the Hertz contact mechanics of the spherical indenter, as follows:

$$\sigma_{33}^{tip}(r) = \begin{cases} -\dfrac{3p}{2\pi a^2}\sqrt{1-\dfrac{r^2}{a^2}}, & r \leq a \\ 0, & r \geq a \end{cases}, \quad (2)$$

where $a = \left(\dfrac{3}{4}\dfrac{pR}{E^*}\right)^{\frac{1}{3}}$ is the contact radius determined by a loading force $p$, a tip radius $R$, and an effective Young's modulus $E^*$. The latter describes the stiffness of the tip-film contact pairing via $1/E^* = (1 - v_{film}^2)/E_{film} + (1 - v_{tip}^2)/E_{tip}$, where $E$ and $v$ are Young's modulus and the Poisson ratio, respectively. At the film-substrate interface, the displacement is continuous for coherency and is assumed to relax to zero within a depth of $h_s$ into the substrate (i.e., $\eta_i|_{x_3=-h_s} = 0$). The clamping effect of the STO substrate is considered to render the average strain zero at each layer of the film (i.e., $\overline{u_{11}} = \overline{u_{22}} = 0$ and $\overline{u_{12}} = 0$). Finally, the boundary value problem of elastic equilibrium, assuming no body force, is given by:

$$\begin{cases} \sigma_{ij,j} = 0 \\ \sigma_{33}|_{x_3=h_f} = \sigma_{33}^{tip}, \;\; \sigma_{31}|_{x_3=h_f} = \sigma_{32}|_{x_3=h_f} = 0 \\ \eta_i|_{x_3=-h_s} = 0 \end{cases}, \quad (3)$$

where stress is related to strain via $\sigma_{ij} = c_{ijkl}e_{kl} = c_{ijkl}(u_{kl} - u_{kl}^0)$. The eigenstrain $u_{ij}^0$ is derived from strain-order parameter couplings of STO through $u_{ij}^0 = Q_{ijkl}P_kP_l + \lambda_{ijkl}q_kq_l$, where $Q_{ijkl}$ and $\lambda_{ijkl}$ are the electrostrictive and rotostrictive tensors, respectively. Using the Khachaturyan microelasticity theory and the Stroh formalism of anisotropic elasticity, we obtained the displacement field of the entire system and then the strain and stress distributions.

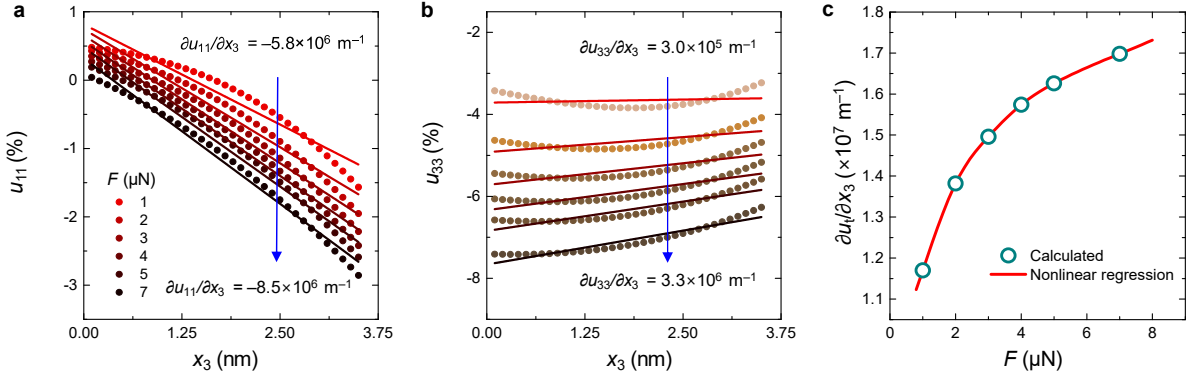

**Supplementary Figure 4. Calculated strain gradients imposed by the AFM tip. a,** The transverse strains ($u_{11}$ values) under the tip (solid circles). The solid lines are the linear fits to the data. The slopes correspond to the transverse strain gradients ($\partial u_{11}/\partial x_3$). **b,** The longitudinal strains ($u_{33}$ values) under the tip (solid circles). The solid lines are the linear fits to the data, and the slopes are the longitudinal strain gradients ($\partial u_{33}/\partial x_3$). Calculations were performed for various applied forces in the range 1–7 μN. **c,** The calculated total transverse strain gradients, $\partial u_t/\partial x_3 = \partial u_{11}/\partial x_3 + \partial u_{22}/\partial x_3$ (open circles). The solid red line indicates the interpolated/extrapolated strain gradients.

Rigorously, the Hertz contact mechanics assumes a non-frictional contact between two isotropic, elastic materials. For dielectric materials, such as the incipient ferroelectric STO, it is generally not linear elastic because of the presence of electromechanical couplings (piezoelectric, flexoelectric, and electrostrictive effects) as well as antiferrodistortive-strain couplings (rotostrictive effect). However, we use the Hertz contact mechanics only to obtain the stress distribution at the STO film surface. With this surface stress distribution as the top boundary condition (and zero displacements at the substrate bottom as the bottom boundary condition), we calculated the stress distribution in the whole system (the film and the substrate) by solving the mechanical equilibrium equation (Supplementary Eq. 3). In our simulation, we also self-consistently take into account the electrostrictive coupling (thereby piezoelectric effects), flexoelectric coupling, and rotostrictive coupling as eigenstrains (stress-free strains). This approach allows us to extend reliably the Hertz contact mechanics to the flexoelectric materials for obtaining stress/strain distribution under the force imparted by the tip.

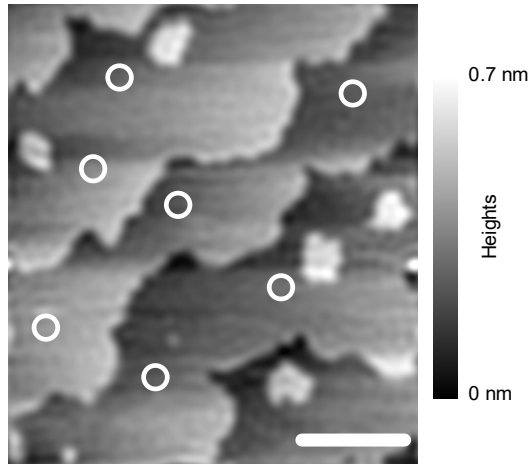

**Supplementary Figure 5. Topography image after tunneling measurements.** Open circles mark the locations where the force dependent tunneling measurements were carried out. This image elaborates that the applied force and bias do not cause either mechanical deformation or electrochemical formation. The scale bar represents 500 nm.

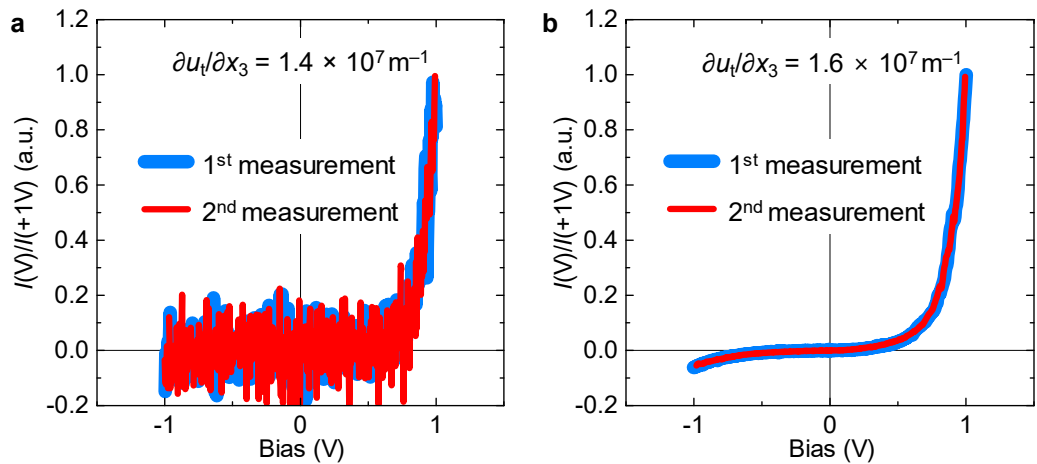

**Supplementary Figure 6. Reversible mechanical control of electron tunneling.** Normalized tunneling currents measured across a nine-unit cell-thick SrTiO$_3$ (STO) film with $\partial u_t/\partial x_3$ of $1.4 \times 10^7$ m$^{-1}$ (**a**) and $1.6 \times 10^7$ m$^{-1}$ (**b**). Blue and red lines indicate the data collected during 1$^{st}$ and 2$^{nd}$ measurements at the same site.

**Supplementary Note 3**

We investigated the effect of strain imposed by the AFM tip. Based on our analytical modeling, the surface stress induced by the AFM tip increased compressive strain by a few % in both the longitudinal and transverse directions. This increase could modify two physical features of the STO layer: (1) the band gap and (2) the physical thickness. First, the band gap of STO increased slightly under compressive strain (decreasing the crystal volume)[3]. Also, according to our strain analysis (Supplementary Fig. 4), pressing by the AFM tip decreased the physical thickness of STO by a few %. We thus incorporated strain-induced systematic changes into the tunnel barrier profiles (Supplementary Figs. 7a,b). However, even after these changes, the $|I_{+1\,V}/I_{-1\,V}|$ values increased only negligibly (Supplementary Fig. 7c). Therefore, any effect of strain per se does not explain our experimental observations.

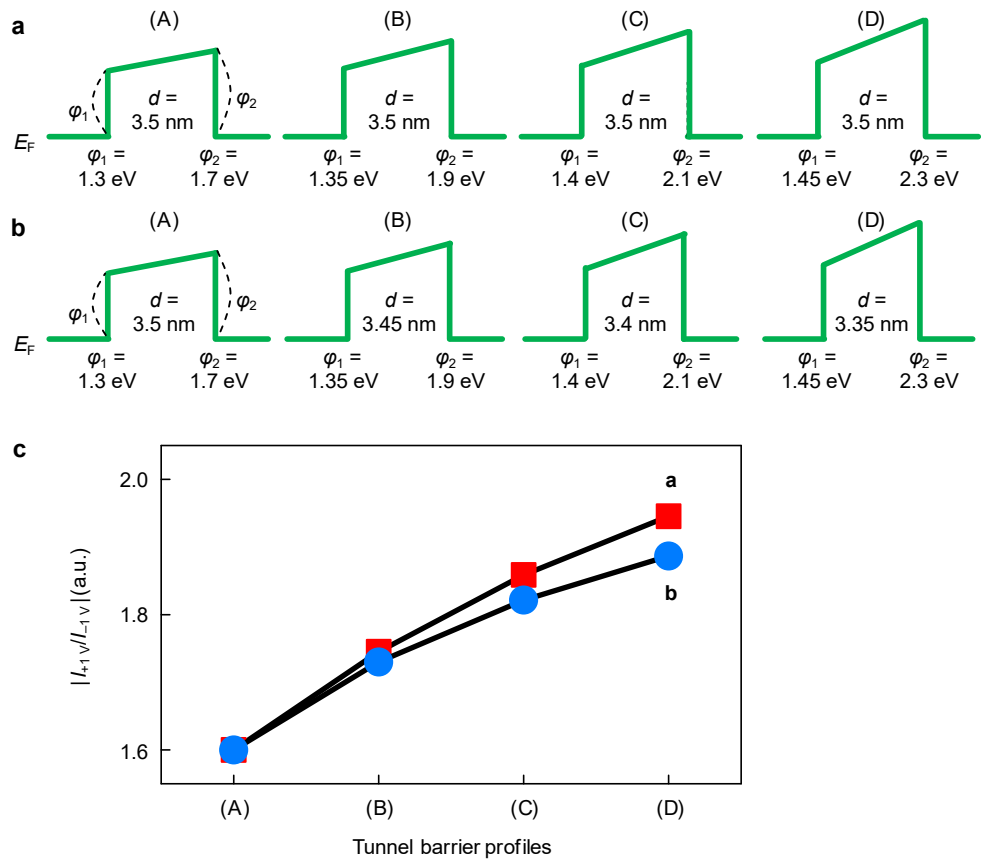

**Supplementary Figure 7. The negligible effect of strain per se. a**, **b**, Systematic modifications of the trapezoidal tunnel barriers. The strains are larger near the STO/PtIr interface than the SRO/STO interface. Thus, any strain-induced increase in the $SrTiO_3$ (STO) band gap would be larger near the STO/PtIr interface, leading to a greater increase in $\varphi_2$ than $\varphi_1$. In **b**, we consider the strain-induced systematic decrease in the barrier width $d$. **c**, *RRs*, (i.e., $|I_{+1\,V}/I_{-1\,V}|$ values) of the tunnel barrier profiles in **a** and **b**, calculated using Supplementary Eq. 1. $\varphi_{1,2}$ represent the barrier heights.

**Supplementary Note 4**

Using the one-dimensional WKB approximation, we can simply describe the tunneling current density for a low $T$ and small $V$, as follows:

$$\begin{aligned} j(V) &= \frac{2e}{h}\int_{-\infty}^{\infty} T(E)\times[f(E)-f(E-eV)]dE \\ &\cong \frac{2e}{h}\int_{-\infty}^{\infty} \exp(-\frac{4\pi}{h}\int_0^d \sqrt{2m(U(x)-E)}dx)\times[f(E)-f(E-eV)]dE, \\ &\cong \frac{2e}{h}\exp(-\frac{4\pi}{h}\int_0^d \sqrt{2m(U(x)-E_F)}dx)\times eV \end{aligned} \quad (4)$$

where $T(E), f(E),$ and $U(x)$ represent the transmission probability, the Fermi-Dirac distribution, and the tunnel barrier profile, respectively.

Using Supplementary Eq. 4, we obtain the tunneling current density for a trapezoidal barrier profile (Supplementary Fig. 8), as follows:

$$\begin{aligned} j(+V) &= \frac{2e}{h}\exp(-\frac{4\pi}{h}\int_0^d \sqrt{2m\left\{\frac{\varphi_2-\varphi_1+eV}{d}(x-d)+\varphi_2\right\}}dx)\times eV \\ &= \frac{2e}{h}\exp(-\frac{8\pi\sqrt{2m}}{3h}\cdot d \cdot \frac{(\varphi_2)^{1.5}-(\varphi_1-eV)^{1.5}}{\varphi_2-\varphi_1+eV})\times eV \end{aligned} \quad (5)$$

$$\begin{aligned} j(-V) &= -\frac{2e}{h}\exp(-\frac{4\pi}{h}\int_0^d \sqrt{2m\left\{\frac{\varphi_2-\varphi_1-eV}{d}(x-d)+\varphi_2-eV\right\}}dx)\times eV \\ &= -\frac{2e}{h}\exp(-\frac{8\pi\sqrt{2m}}{3h}\cdot d \cdot \frac{(\varphi_2-eV)^{1.5}-(\varphi_1)^{1.5}}{\varphi_2-\varphi_1-eV})\times eV \end{aligned} \quad (6)$$

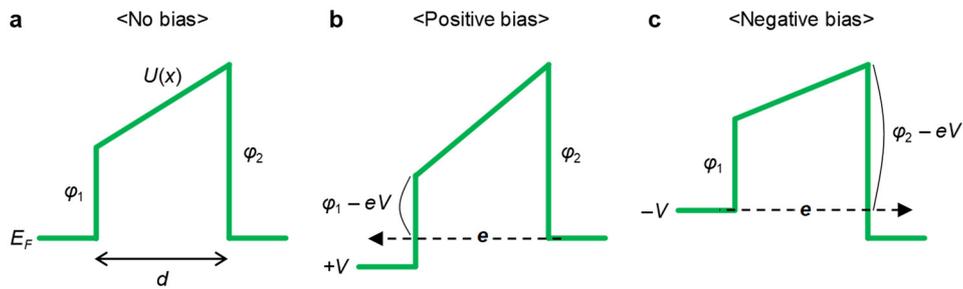

**Supplementary Figure 8. Trapezoidal tunnel barriers.** Schematics of a trapezoidal tunnel barrier under no bias (**a**) and under positive (**b**) and negative (**c**) bias. $\varphi_{1,2}$ represent the barrier heights.

Using Supplementary Eq. 4, we obtain the tunneling current density for a triangular barrier profile (Supplementary Fig. 9), as follows:

$$j(+V) = \frac{2e}{h}\exp\left(-\frac{4\pi}{h}\int_0^{d'}\sqrt{2m\left\{\frac{\varphi}{d'}(x-d')+\varphi\right\}}dx\right)\times eV$$
$$= \frac{2e}{h}\exp\left(-\frac{8\pi\sqrt{2m}}{3h}\cdot d \cdot \frac{\varphi^{1.5}}{\varphi+eV}\right)\times eV \quad (7)$$

$$j(-V) = -\frac{2e}{h}\exp\left(-\frac{4\pi}{h}\int_0^{d}\sqrt{2m\left\{\frac{\varphi-eV}{d}(x-d)+\varphi-eV\right\}}dx\right)\times eV$$
$$= -\frac{2e}{h}\exp\left(-\frac{8\pi\sqrt{2m}}{3h}\cdot d \cdot (\varphi-eV)^{0.5}\right)\times eV \quad (8)$$

Finally, using Supplementary Eqs. (5–8), we obtain tunneling *I–V* curves for systematically modified tunnel barrier profiles (Supplementary Fig. 10). In these calculations, we assume $d = 3.5$ nm.

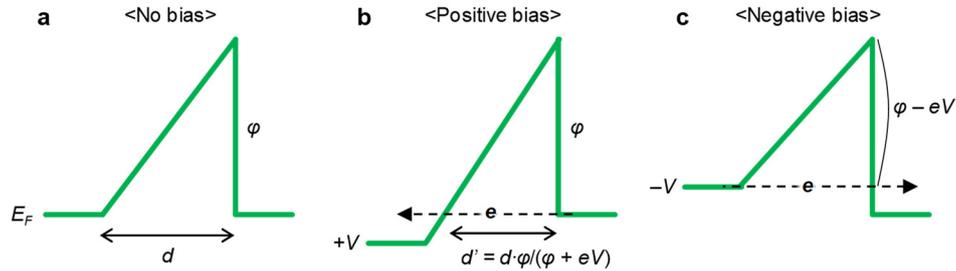

**Supplementary Figure 9. Triangular tunnel barriers.** Schematics of a triangular tunnel barrier under no bias (**a**) and under positive (**b**) and negative (**c**) bias. $\varphi$ represents the barrier height.

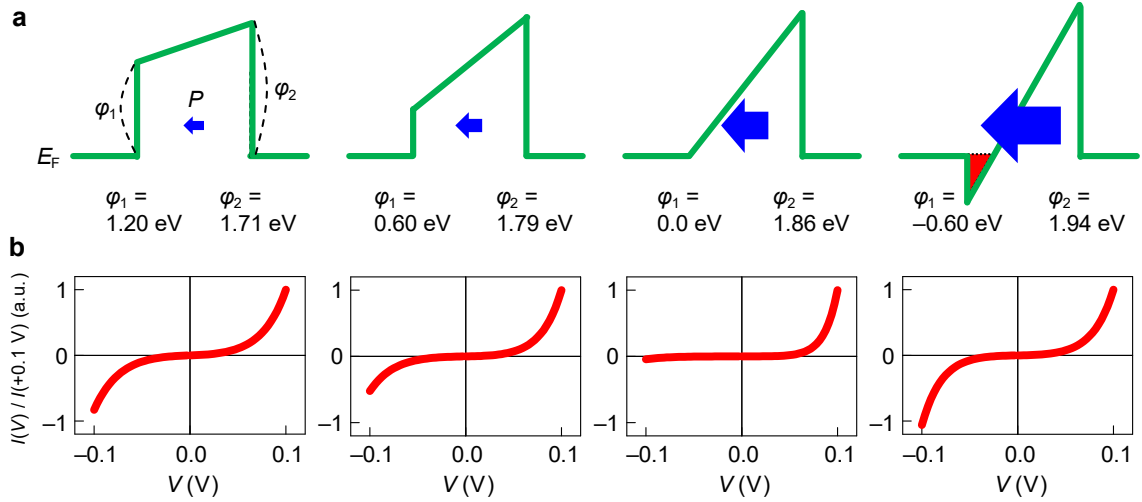

**Supplementary Figure 10. Systematic changes in tunnel currents.** Tunnel barrier profiles (**a**) and the corresponding tunneling *I–V* curves (**b**), calculated using the Wentzel–Kramers–Brillouin (WKB) approximation. $\varphi_{1,2}$ represent the barrier heights.

**Supplementary Note 5**

To understand the effect of the electronic polarization to the interfacial electronic structure, we constructed SRO/STO bilayer with 5 unit cells of SRO and 9 layer of STO, part of which is shown in Fig. 3a. The sub-interfacial layers of completely relaxed paraelectric phase of STO on SRO are insulating as can be seen from the plot of layer resolved density of states (LDOS) by black continuous lines as shown in Fig. 3b, in which Fermi level lies in the gap between the conduction band minima and valance band maxima. However, when STO is polarized, the induced field bends bands and brings the bottom of conduction bands of sub-interfacial STO layers below Fermi level, as shown by blue filled curves in the plot of LDOS (Fig. 3b). We plot Fig. 3 with frozen uniform displacement of Ti atom by 0.2 Å. Note that polarized tetragonal STO have higher energy than paraelectric cubic STO, but can be stabilized in non-equilibrium strain conditions[4]. This band profile clearly supports the experimental finding that the metallized interfacial STO layer changes the barrier profile from trapezoidal to triangular.

**Supplementary Note 6**

To understand how the strain affects the band structure of SrRuO3 (SRO) and subsequently the tunnelling transport, we additionally performed first-principles DFT calculations. We fixed the in-plane lattice parameter of SRO to that of STO substrate, and imposed compressive strain $u_{33}$ (ranging from 0 to –8%) in the out-of-plane direction. This assumption closely accounts for the strain distribution, obtained from the phase-field simulations (Supplementary Fig. 4). As shown in the Supplementary Figs. 11a and b, our calculation suggests that with increasing the strain, the density of states at the Fermi energy ($\rho_F$) slightly increases and thus the screening length ($\delta_{SRO} \propto 1/\sqrt{\rho_F}$) could decrease, whereas the work function of SRO ($W_{SRO}$) slightly decreases by ~0.2 eV. Given the electrostatic constraint $\Delta\varphi_1/\Delta\varphi_2 = (\varphi_{0,1} - \varphi_1)/(\varphi_2 - \varphi_{0,2}) = \delta_{SRO}/\delta_{PtIr}$, where $\varphi_{0,1}$ is proportional to $W_{SRO}$, the influence of the decreased $\delta_{SRO}$ on the tunnel barriers seems to cancel out that of the decreased $W_{SRO}$. Furthermore, these changes in $\delta_{SRO}$ and $W_{SRO}$ are too small to be responsible for the anomalous behavior of tunnel transports (Supplementary Fig. 11c). Thus, we conclude that the effect of strain on SRO is not significant.

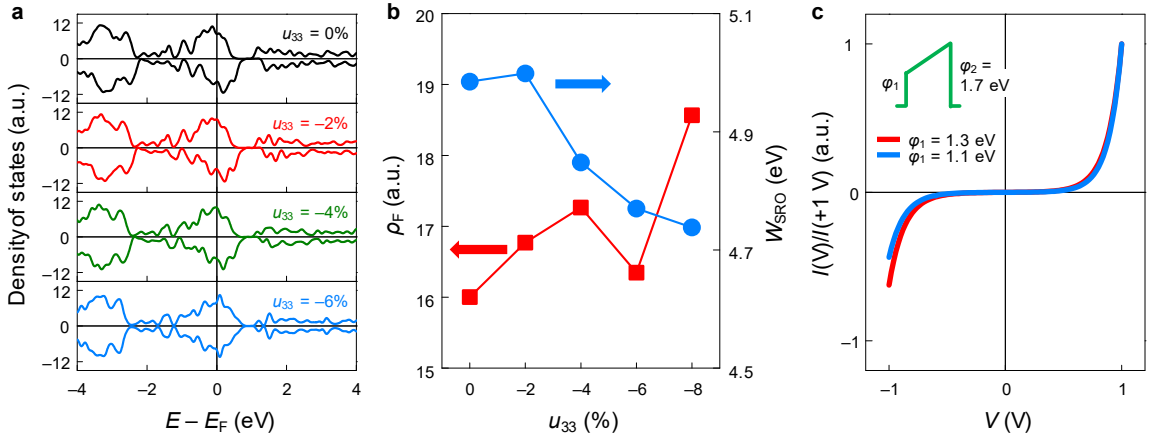

**Supplementary Figure 11. Effect of strain on the SrRuO3 electrode. a**, Density of states of SrRuO3 (SRO) for different out-of-plane strains $u_{33}$. **b**, Density of states at the Fermi energy ($\rho_F$; red squares) and work function ($W_{SRO}$; blue circles) of SRO as a function of $u_{33}$. **c**, Tunneling *I–V* curves, corresponding to the tunnel barrier profiles (inset), calculated using Supplementary Eq. 1.

**Supplementary Note 7**

For an ultrathin polarized STO layer sandwiched by SRO and PtIr metals, free carriers in SRO and PtIr partly screen the surface charges in the polarized STO layer. Considering the screening charge, the flexoelectric polarization should modify the tunnel barrier profile according to the following electrostatic equations[5,6]:

$$(\varphi_2 - \varphi_1)/ed = (P - \sigma_s)/\varepsilon + E_{bi} = [f_{eff} \cdot (\partial u_t/\partial x_3) - \sigma_s/\varepsilon] + E_{bi}, \quad (9)$$

$$\sigma_s = d \cdot P/[(\varepsilon/\varepsilon_0) \cdot (\delta_{SRO} + \delta_{PtIr}) + d], \quad (10)$$

where $\sigma_s$ is the magnitude of the screening charge per unit area. Given $\varepsilon/\varepsilon_0 = 40$, $\delta_{SRO} + \delta_{PtIr} = 0.7$ nm, and $d = 3.5$ nm, Supplementary Eq. 10 estimates $\sigma_s$ to be as small as $0.11P$. For an order-of-magnitude estimate of $f_{eff}$ in the main text, therefore, we just used the simplified electrostatic equation 1 in the main text, where we neglect the effect of $\sigma_s$. If we additionally consider the effect of $\sigma_s$, the linear slope (i.e., $23 \pm 1$ V) in the $(\varphi_2 - \varphi_1)/ed$ vs. $\partial u_t/\partial x_3$ curve (Fig. 4b) corresponds to $(1 - 0.11)f_{eff} = 0.89 f_{eff}$. Thus, this correction gives $f_{eff} = (23/0.89)$ V = 26 V.

**Supplementary Note 8**

At the critical polarization, $P_c$, we expect that $\varphi_1 = 0$ and $\varphi_2 = \varphi_{0,2} + \varphi_{0,1} \cdot (\delta_{PtIr}/\delta_{SRO}) = 1.7 + 1.3/8$ eV, yielding $(\varphi_2 - \varphi_1)_c/ed = 5.32 \times 10^8$ V m$^{-1}$. With $E_{bi} = 9 \times 10^7$ V m$^{-1}$, obtained from fitting (Fig. 3b), we can roughly estimate the critical polarization as $P_c = \varepsilon \cdot [(\varphi_2 - \varphi_1)_c/ed - E_{bi}] \sim 0.156$ C m$^{-2}$, where we used $\varepsilon = 40\varepsilon_0$ based on the average strain state (i.e., $u_{33} = -0.06$ and $u_{11} = u_{22} = -0.01$; Supplementary Fig. 4) and DFT calculation (Supplementary Fig. 12).

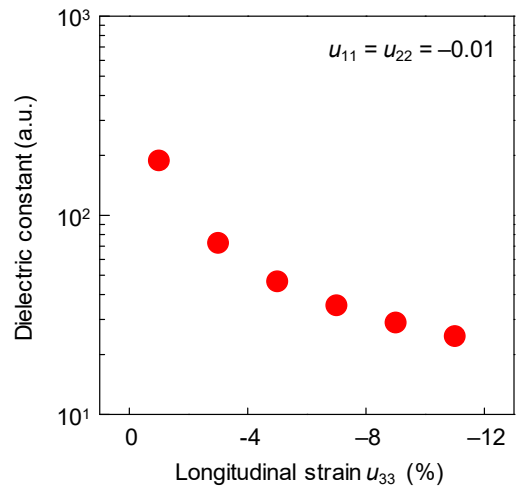

**Supplementary Figure 12. Calculated dielectric constant as a function of longitudinal strain.** Density functional theory (DFT) calculations of the out-of-plane component of the total dielectric constant (i.e., $\varepsilon_{zz}$), which includes both ionic and electronic contributions, as a function of strain $u$. The strain was measured with respect to the DFT equilibrium lattice of 3.86 Å.

**Supplementary Note 9**

A nonlinear flexoelectric response could arise under large strain gradients, as demonstrated in several material systems[7,8]. In the case of a centrosymmetric material like STO, the quadratic flexoelectric term should be zero, so we additionally considered the cubic flexoelectric term, i.e., $P/\varepsilon = f \cdot (\partial u_t/\partial x_3) + g \cdot (\partial u_t/\partial x_3)^3$, where $f$ and $g$ are the first-order and third-order flexocoupling coefficients. For simplicity, by assuming $f = 2.6$ V (i.e., bulk flexocoupling coefficient)[9], we fitted our data and found that $g$ is minuscule, as small as $3.8 \times 10^{-14}$ V m$^2$ (Supplementary Fig. 13). However, when $\partial u_t/\partial x_3$ is huge, e.g., much larger than $10^5$ m$^{-1}$, the effective flexocoupling coefficient, i.e., $f_{\text{eff}} = (P/\varepsilon)/(\partial u_t/\partial x_3) = f + g \cdot (\partial u_t/\partial x_3)^2$, might become significantly enhanced.

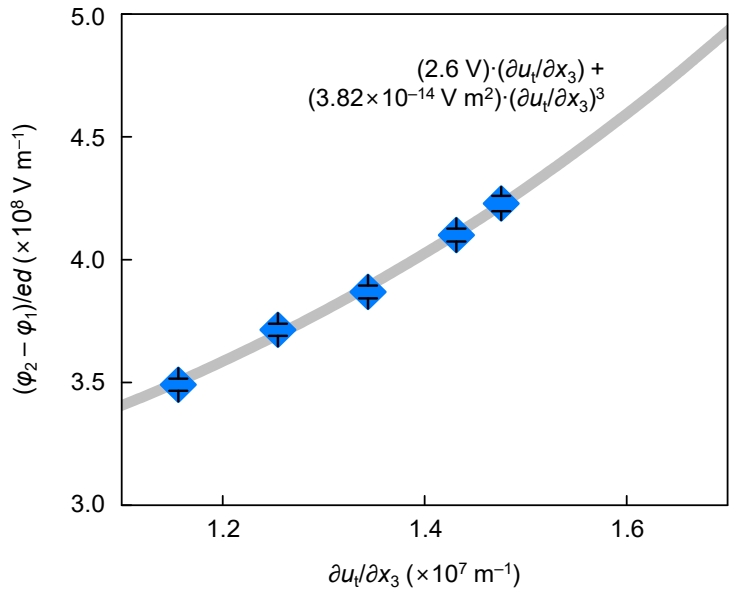

**Supplementary Figure 13. Analysis under the assumption of the third-order flexoelectricity**. Filled symbols are calculated from Fig. 4a in the main text. The error bars represent the standard deviations obtained by fitting the tunneling spectra in the Supplementary Figs. 2 a-e to the Supplementary Eq. 1. The gray line shows a fit to $f \cdot (\partial u_t/\partial x_3) + g \cdot (\partial u_t/\partial x_3)^3$ with assuming $f = 2.6$ V.

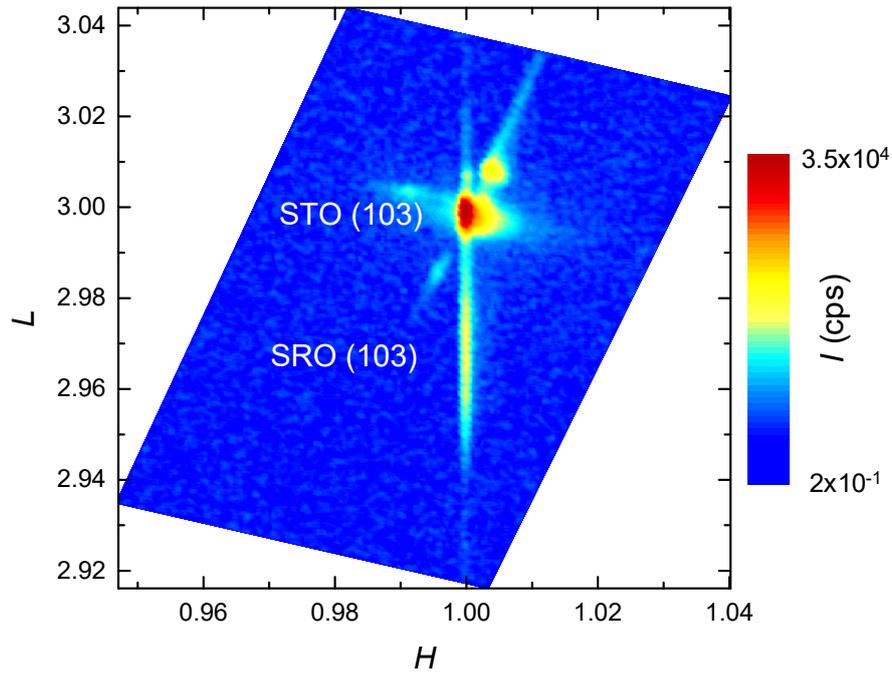

**Supplementary Figure 14. Structural characterization of the nine-unit cell thick SrTiO$_3$ film.** Reciprocal space mapping around SrTiO$_3$ (STO) (103) Bragg reflex. Except the peaks from the STO substrate and bottom SrRuO$_3$ (SRO) layer, we did not detect any additional Bragg peak from the STO film. Therefore, this data suggests that our ultrathin STO barrier layer is strain-free.

**Supplementary References**